\documentclass[aps,preprint,groupedaddress,showpacs,prd,epsf]{revtex4}
\usepackage{graphics}
\begin{document}

\thispagestyle{empty}
{\baselineskip0pt
\leftline{\large\baselineskip16pt\sl\vbox to0pt{\hbox{\it Department of
Physics}
               \hbox{\it Osaka City  University}\vss}}
\rightline{\large\baselineskip16pt\rm\vbox to20pt{\hbox{OCU-PHYS-199}
            \hbox{AP-GR-9}\hbox{YITP-03-18}
\vss}}%
}
\vskip3cm

\title{High Speed Dynamics of Collapsing Cylindrical Dust Fluid}


\author{Ken-ichi Nakao${}^{1}$ and Yoshiyuki Morisawa${}^{1,2}$}
\affiliation{${}^{1}$Department of Physics, Graduate School of Science,
Osaka City University, Osaka 558-8585, Japan \\
${}^{2}$Yukawa Institute for Theoretical Physics, 
Kyoto University, Kyoto 606-8502, Japan}


\date{\today}

\begin{abstract}
We construct approximate solutions that will describe the
last stage of cylindrically symmetric gravitational collapse of dust
fluid. Just before the spacetime singularity formation,
the speed of the dust fluid might be almost equal to the speed of
light by gravitational acceleration. Therefore the analytic solution
describing the dynamics of 
cylindrical null dust might be the crudest approximate 
solution of the last stage of the gravitational collapse.
In this paper, we regard this null dust solution as a background and
perform `high-speed approximation' to know the gravitational collapse 
of ordinary timelike dust fluid; the `deviation of the timelike
 4-velocity vector field from null' is treated as a perturbation.
In contrast with the null dust approximation, our approximation scheme can
describe the generation of gravitational waves in the course of the
cylindrically symmetric dust collapse.

\end{abstract}


\maketitle

\section{Introduction}

The gravitational collapse and spacetime singularity formation are 
physically very significant phenomena predicted by general
relativity. However since extremely high density, high pressure and large spacetime
curvature will be realized around the spacetime singularity,
all known theories of physics including general relativity could break down
there and a quantum theory of gravity might be necessary to describe 
such physical situations. 
If this perspective is right, the appearance of the spacetime
singularity predicted by general relativity has no rigorous meaning,  
since general relativity can not describe the 
gravitational processes in the neighborhood of the spacetime singularity. 
However, it is still important to study the spacetime structures 
around singularities by assuming general relativity, 
since such studies reveal what happens at the entrance of new
physics, although there is an issue of their 
observability in connection with the cosmic censorship\cite{Ref:penrose69}.

A singularity visible for distant observers  
is called a globally naked singularity. 
Nakamura, Shibata and one of the
present authors (NSN) have conjectured that the
large spacetime curvature in the neighborhood of non-spherical spacetime
singularities can propagate away to infinity in the form of gravitational
radiation if those singularities are globally naked\cite{Ref:DRIES-UP}.
If this conjecture is true, formation processes of naked singularities
are strong sources of gravitational radiation and thus 
might be one of the targets of gravitational wave 
astronomy\cite{Ref:TAMA,Ref:LIGO,Ref:VIRGO,Ref:GEO,Ref:LISA}. 
However, numerical simulations performed by Shapiro and Teukolsky 
suggested that the gravitational collapse of a 
collisionless gas spheroid might form a spindle 
naked singularity in accordance with 
the hoop conjecture\cite{Ref:HOOP,Ref:hoop-misc}, 
but little gravitational radiation is generated 
in its formation process\cite{Ref:Spindle}. 
This seems to be a counter example against NSN's conjecture.  
However it should be noted that the numerical accuracy of 
the simulations performed by Shapiro and Teukolsky seems to be 
insufficient to give a definite statement\cite{Ref:DRIES-UP}. 

Motivated by the work of Shapiro and Teukolsky, 
Echeverria performed numerical simulations with very high resolution  
for the cylindrically symmetric gravitational collapse of an 
infinitely thin dust shell\cite{Ref:Dust-Shell}. 
Although infinitely long cylindrical matter or radiation
is unrealistic, this system could be a crude approximation of
sufficiently thin and long but finite length distribution
like as the case of spindle gravitational collapse 
simulated by Shapiro and Teukolsky.  
As Thorne showed, the gravitational collapse of 
a cylindrical matter with a compact support in 
the radial direction does not form a black hole horizon
(marginal surface)\cite{Ref:HOOP}. Thus the singularities formed 
in the simulations by Echeverria must be globally naked. 
Further Echeverria's numerical results showed that the gravitational 
collapse of the cylindrical dust shell causes gravitational wave burst. 
Thus in this case, the naked singularity formation is really a 
strong source of gravitational radiation. 
This seems to be inconsistent with Shapiro and Teukolsky. 

By the same motivation as that of Echeverria, Chiba  
investigated the gravitational collapse 
of cylindrically distributed dust fluid, which also forms a naked
singularity. Then Chiba's numerical results imply
that the gravitational radiation generated by
dust collapse is extremely small\cite{Ref:Chiba}.
This is consistent with the results by Shapiro and Teukolsky and further 
with Piran's numerical simulations for the gravitational collapse 
of cylindrical perfect fluid with the adiabatic index
$4/3\leq\Gamma\leq2$; softer equation of state leads to less emission 
of gravitational waves\cite{Ref:Piran}. 

In this paper, in order to clarify the reason of the 
apparent inconsistency between Echeverria's work and the others, 
and further to understand the generation 
mechanism of gravitational radiation in the formation of linear 
singularities, we investigate the dynamics of the cylindrically symmetric
dust fluid by the analytic procedure. 
Here we should note that it is rather dangerous to accept  
numerical results, especially these for structures 
around spacetime singularities, without any consistency checks. 
In this sense, it is also of great significance 
to obtain approximate solutions by completely or almost analytic procedures, 
which could complement the numerical studies. 

A few exact solutions for cylindrical gravitational collapse are known. 
One is that describing the gravitational collapse of null 
dust\cite{Ref:Morgan,Ref:LW,Ref:Nolan}, which 
is a cylindrically symmetric version of Vaidya
solution\cite{Ref:DJ89}. 
As gravitational collapse proceeds, 
the speed of collapsing matter might approach to the speed of light. 
Thus the null dust is often regarded 
as the crudest approximation of the last stage of gravitational 
collapse. However the cylindrical null dust 
does not generate gravitational radiation. 
Recently, Pereira and Wang constructed analytic model 
of the cylindrical thin shell made of counter 
rotating dust particles\cite{Ref:PW}. 
This exact solution might be useful and 
physically significant, but unfortunately, it is impossible to describe
the generation of gravitational radiation near the spacetime
singularity\cite{Ref:GJ}. 
Here we would like to comment on Echeverria's analytic work. 
Echeverria analytically constructed an asymptotic solution 
for the gravitational collapse of an infinitely thin dust shell 
by using the numerical results and further by putting self-similar 
ansatz\cite{Ref:Dust-Shell}. 
Although Echeverria's asymptotic solution might describe the generation 
of gravitational waves, the behavior of the infinitely thin shell 
might be different from that of finitely distributed masses. 
Therefore in order to see the gravitational wave generation 
in more general situations analytically, 
we need new exact solutions or new approximation scheme. 

In this paper, we perform a kind of perturbation analysis 
on the null dust solution by assuming that the speed of the 
collapsing dust is almost equal to the speed of light. 
We stress that the generation of gravitational 
radiation can be described by this approximation scheme. 

This paper is organized as follows. In Sec.II, we present
the basic equations of the cylindrically symmetric dust system.
In Sec.III, we derive the equations and solutions of
our approximation scheme to describe the gravitational collapse
with almost the speed of light; 
we explicitly show solutions for thin and thick
dust shells, separately. 
Finally, Sec.IV is devoted for summary and discussion of the
physical meaning of our solutions by comparing those with the
previous studies.

In this paper, we adopt $c=1$ unit.

\section{Cylindrically Symmetric Dust System}

In this paper, we assume the whole-cylinder symmetry which leads to  
the following line 
element\cite{Ref:MelvinI,Ref:C-energy,Ref:MelvinII},
\begin{equation}
ds^{2}=e^{2(\gamma-\psi)}\left(-dt^{2}+dr^{2}\right)
+e^{2\psi}dz^{2}+e^{-2\psi}R^{2}d\varphi^{2},
\end{equation}
where $\gamma$, $\psi$ and $R$ are functions of $t$ and $r$.
Then Einstein equations are
\begin{eqnarray}
&&\gamma'=\left({R'}^{2}-{\dot R}^{2}\right)^{-1}
\biggl\{
RR'\left({\dot \psi}^{2}+{\psi'}^{2}\right)
-2R{\dot R}{\dot \psi}\psi'
+R'R''-{\dot R}{\dot R}' \nonumber \\ 
&&~~~~-8\pi G\sqrt{-g}\left(R'T^{t}{}_{t}+{\dot R}T^{r}{}_{t}\right)
\biggr\}, \\
&&{\dot \gamma}=-\left({R'}^{2}-{\dot R}^{2}\right)^{-1}
\biggl\{
R{\dot R}\left({\dot \psi}^{2}+{\psi'}^{2}\right)
-2RR'{\dot \psi}\psi'
+{\dot R}R''-R'{\dot R}' \nonumber \\
&&~~~~-8\pi G\sqrt{-g}\left({\dot R}T^{t}{}_{t}+R'T^{r}{}_{t}\right)
\biggr\}, \\
&&{\ddot \gamma}-\gamma''={\psi'}^{2}-{\dot \psi}^{2}-{8\pi G\over R}
\sqrt{-g}T^{\varphi}{}_{\varphi}, \\
&&{\ddot R}-R''
=-8\pi G\sqrt{-g}\left(T^{t}{}_{t}+T^{r}{}_{r}\right), \\
&&{\ddot \psi}+{{\dot R}\over R}{\dot \psi}-\psi''
-{R'\over R}\psi'
=-{4\pi G\over R}\sqrt{-g}\left(T^{t}{}_{t}+T^{r}{}_{r}-T^{z}{}_{z}
+T^{\varphi}{}_{\varphi}\right),
\end{eqnarray}
where $g$ is the determinant of the metric tensor, 
a dot means the derivative with respect to $t$ while a prime means
the derivative with respect to $r$.

In this paper, we consider dust fluid. 
The stress-energy tensor is
\begin{equation}
T^{\mu\nu}=\rho u^{\mu}u^{\nu},
\end{equation}
where $\rho$ is the rest mass density and $u^{\mu}$ is the 4-velocity
field; all these matter variables depend on 
$t$ and $r$ only. 
By the assumption of the whole-cylinder symmetry,
the components of the 4-velocity $u^{\mu}$ are written as
\begin{equation}
u^{\mu}=u^{t}\left(1,-1+V,~0,~0\right).
\end{equation}
By the normalization $u^{\mu}u_{\mu}=-1$, $u^{t}$ is expressed as
\begin{equation}
u^{t}={e^{-\gamma+\psi}\over \sqrt{V\left(2-V\right)}}.
\end{equation}

Here we introduce a new density variable $D$ defined by
\begin{equation}
D:={\sqrt{-g}\rho u^{t}\over \sqrt{V\left(2-V\right)}}
={R e^{\gamma-\psi}\rho \over V\left(2-V\right)}.
\label{eq:D-def}
\end{equation}
The components of the stress-energy tensor are then expressed as
\begin{eqnarray}
\sqrt{-g}T^{t}{}_{t}&=&-e^{\gamma-\psi}D, \\
\sqrt{-g}T^{r}{}_{t}&=&e^{\gamma-\psi}D(1-V)=-\sqrt{-g}T^{t}{}_{r}, \\
\sqrt{-g}T^{r}{}_{r}&=&e^{\gamma-\psi}D(1-V)^{2},
\end{eqnarray}
and the other components vanish.
The equation of motion $\nabla_{\beta}T^{\beta}{}_{\alpha}=0$ 
($\nabla_{\beta}$ is the covariant derivative) leads to
\begin{eqnarray}
\partial_{u}D&=&-{1\over2}(DV)'+{D\over2}(1-V)
  \left\{2\partial_{u}\left(\psi-\gamma\right)-V\left({\dot \psi}
  -{\dot \gamma}\right)\right\}, \label{eq:conservation-full} \\
D\partial_{u}V&=&(1-V)\partial_{u}D
  +{1\over2}\left\{V\left(1-V\right)D\right\}
  -{D\over2}\left\{2\partial_{u}\left(\psi-\gamma\right)
  -V({\dot \psi}-{\dot\gamma})\right\}, \label{eq:Euler-full}
\end{eqnarray}
where $u=t-r$ is the retarded time and $\partial_{u}$ is the partial
derivative of $u$ with the advanced time $v=t+r$ fixed.

To estimate the energy released in the form of gravitational radiation,
the $C$-energy and its flux vector proposed by Thorne\cite{Ref:C-energy} is
useful. The $C$-energy $E$ is given as
\begin{equation}
E:={1\over8}\left\{1-e^{-2\psi}\left(\nabla^{\alpha}R\right)
\left(\nabla_{\alpha}R\right)\right\}.
\end{equation}
Then the flux vector $J^{\mu}$ associated with $E$ is introduced as
\begin{equation}
\sqrt{-g}J^{\mu}={1\over 2\pi}(\partial_{r}E,~\partial_{t}E,~0,~0).
\end{equation}
As shown by Hayward, if the null energy condition is satisfied, 
$\zeta^{\alpha}\nabla_{\alpha}E\geq0$ for any achronal vector 
$\zeta^{\alpha}$, and hence $0\leq E <1/8$ holds outside the 
causal future of the naked singularity\cite{Ref:Hayward}.
Note that the conservation law $\nabla_{\mu}J^{\mu}=0$ is trivially
satisfied. Using Einstein equations, the energy flux vector $J^{\mu}$ is
expressed as
\begin{eqnarray}
\sqrt{-g}J^{t}&=&{e^{-2\gamma}\over 8\pi G}
\biggl\{RR'({\dot \psi}^{2}+{\psi'}^{2})
-2R{\dot R}{\dot \psi}\psi'
-8\pi G\sqrt{-g}(R'T^{t}{}_{t}+{\dot R}T^{r}{}_{t})\biggr\}, \\
\sqrt{-g}J^{r}&=&{e^{-2\gamma}\over 8\pi G}
\biggl\{R{\dot R}({\dot \psi}^{2}+{\psi'}^{2})
-2RR'{\dot \psi}\psi'
-8\pi G\sqrt{-g}(R'T^{r}{}_{t}-{\dot R}T^{r}{}_{r})\biggr\},
\end{eqnarray}
and the other components vanish.

\section{The Gravitational Collapse of High Speed Dust}

Echeverria showed that the speed of a collapsing cylindrical dust 
shell becomes faster and faster and finally almost equal to 
the speed of light. This seems to be a generic behavior  
near the spacetime singularity even in more general situations, and 
thus hereafter we focus on the situation in which speeds of dust fluid 
elements are almost equal to the speed of light.
On the ground of physics, such a situation 
can be well approximated by Morgan's null dust solution\cite{Ref:Morgan}. 
Then we treat the `deviation of the 4-velocity of the dust fluid
from null' as a perturbation and will perform a linear perturbation
analysis. 

First, we consider the null dust limit of the timelike dust fluid. 
The stress-energy tensor is rewritten as
\begin{equation}
T_{\mu\nu}={e^{-3(\gamma-\psi)}D\over R} k_{\mu}k_{\nu},
\end{equation}
where
\begin{equation}
k^{\mu}=(1,-1+V,~0,~0).
\end{equation}
In the limit of $V\rightarrow0$, $k^{\mu}$ approaches
to the null vector and hence $T_{\mu\nu}$ agrees with that of
null dust if this limit is taken with $D$ fixed.
Here note that the limit of $V\rightarrow0$ with $D$ fixed leads to 
vanishing rest mass density $\rho\rightarrow0$, i.e., massless field   
(see Eq.(\ref{eq:D-def})). 
Thus for $0<V\ll1$, in order to obtain the approximate 
solutions for the metric variables, the stress-energy tensor could 
be replaced by that of collapsing null dust. 

Assuming that $\psi$ vanishes initially,
the solutions of the complete null dust are easily obtained as
\begin{eqnarray}
\psi&=&0, \\
\gamma&=&\gamma_{B}(v), \label{eq:gamma-BG}\\
R&=&r, \\
8\pi G De^{\gamma_{B}}&=&{d\gamma_{B}\over dv}, \label{eq:D-BG}
\end{eqnarray}
where $\gamma_{B}(v)$ is an arbitrary function of the
advanced time $v$\cite{Ref:Morgan}.
This solution was studied by Letelier and Wang\cite{Ref:LW} and Nolan in
detail\cite{Ref:Nolan}.
If the density variable $D$ does not vanish at $r=0$, there is the 
naked singularity; this is not a scalar polynomial singularity 
but intermediate one at which observers suffer infinite tidal force 
although all the scalar polynomials of the Riemann tensor do not 
diverge\cite{Ref:LW}. 
The situation is depicted in Fig.1. In this figure, we assume that 
the density variable $D$ has a compact support with respect to 
the advanced time $v$, i.e., $D>0$ for $0<v<v_{\rm w}$. 
As mentioned, we will regard this solution 
as a background spacetime. 

We introduce a small parameter $\epsilon$ and assume the order of
the variables as $V=O(\epsilon)$ and $\psi=O(\epsilon)$.
Further we rewrite the variables $\gamma$, $R$ and $D$ as
\begin{eqnarray}
e^{\gamma}&=&e^{\gamma_{B}}(1+\delta_{\gamma}), \\
R&=&r(1+\delta_{R}), \\
D&=&D_{B}(1+\delta_{D}),
\end{eqnarray}
and assume that $\delta_{\gamma}$, $\delta_{R}$
and $\delta_{D}$ are also $O(\epsilon)$, where
\begin{equation}
D_{B}(v):={1\over 8\pi Ge^{\gamma_{B}}}{d\gamma_{B}\over dv}.
\label{eq:DB-def}
\end{equation}
We would like to call the perturbative analysis with respect to this small
parameter $\epsilon$ `the high-speed approximation'.

The 1st order equations with respect to $\epsilon$ are
given as follows; the Einstein equations lead to
\begin{eqnarray}
&&\delta_{\gamma}{}'=
8\pi Ge^{\gamma_{B}}D_{B}\left\{\delta_{\gamma}-\psi+\delta_{D}
-2\partial_{v}(r\delta_{R})\right\}
+(r\delta_{R})'',\label{eq:del-r-gamma}\\
&&{\dot \delta}_{\gamma}=
8\pi Ge^{\gamma_{B}}D_{B}\left\{\delta_{\gamma}-\psi+\delta_{D}
-2\partial_{v}(r\delta_{R})-V\right\}
+(r{\dot\delta_{R}})', \label{eq:del-t-gamma}\\
&&{\ddot\delta}_{\gamma}-\delta_{\gamma}{}''=0, \label{eq:gamma-linear}\\
&&r{\ddot\delta}_{R}-(r\delta_{R})''=16\pi Ge^{\gamma_{B}}D_{B}V,
\label{eq:R-linear}\\
&&{\ddot\psi}-\psi''-{1\over r}\psi'={8\pi G \over r}e^{\gamma_{B}}D_{B}V,
\label{eq:psi-linear}
\end{eqnarray}
where $\partial_{v}$ is the partial
derivative of the advanced time $v$ with fixed retarded time $u$;
the conservation law (\ref{eq:conservation-full}) leads to
\begin{equation}
\partial_{u}\left(\delta_{D}+\delta_{\gamma}-\psi\right)
=-{e^{\gamma_{B}}\over2D_{B}}
\left\{V{d\over dv}\left(e^{-\gamma_{B}}D_{B}\right)
+V'e^{-\gamma_{B}}D_{B}\right\};
\label{eq:conservation-linear}
\end{equation}
the Euler equation (\ref{eq:Euler-full}) reads as
\begin{equation}
\partial_{u}V=0,  \label{eq:V-linear}
\end{equation}
where we have used Eq.(\ref{eq:conservation-full}).
Equation(\ref{eq:V-linear}) means that $V$ is an arbitrary function of
the advanced time $v$, i.e., 
\begin{equation}
V=V(v).  \label{eq:V-sol}
\end{equation}
This implies that $V$ does not vanish
even in the limit to the spacetime singularity. 

Since we solve the basic equations 
(\ref{eq:del-r-gamma})-(\ref{eq:V-linear}) as a Cauchy problem, 
we have to set initial data for these equations. 
Our purpose is to study the generation mechanism of gravitational 
radiation in the formation process of the naked singularity. 
Therefore we set almost all the perturbation variables to vanish at  
some moment before the formation of the naked singularity. 
More concretely speaking, we set the initial data as 
\begin{equation}
\delta_{\gamma}=\delta_{R}={\dot\delta}_{R}
=\psi={\dot \psi}=\delta_{D}=0~~~~~~{\rm at}~~t=t_{\rm i},
\label{eq:initial}
\end{equation}
where $t_{\rm i}$ is chosen so that
the spacetime singularity of the background 
does not form before and at $t=t_{\rm i}$.
Then we may recognize the formation process of the singularity 
as an origin of the perturbation variables generated after $t=t_{\rm i}$. 
Here note that these initial conditions
are consistent with Eq.(\ref{eq:del-r-gamma}).

It should be noted that $V$ and ${\dot\delta}_{\gamma}$ 
have to have non-trivial values at $t=t_{\rm i}$. 
Since we consider the timelike dust, $V$ must not vanish. 
Due to this constraint and Eq.(\ref{eq:del-t-gamma}), 
${\dot\delta}_{\gamma}$ has non-trivial value at $t=t_{\rm i}$ as 
\begin{equation}
{\dot\delta}_{\gamma}(t_{\rm i},r)=-S(t_{\rm i}+r),
\label{eq:gamma-initial}
\end{equation}
where for notational simplicity, we have introduced a new variable 
$S$ defined by
\begin{equation}
S(v):=8\pi G e^{\gamma_{B}}D_{B}V.
\end{equation}

For $t\geq t_{\rm i}$,
the solution of $\delta_{\gamma}$ satisfying the initial conditions
(\ref{eq:initial}) and (\ref{eq:gamma-initial}) is obtained as
\begin{equation}
\delta_{\gamma}(t,r;t_{\rm i})=
{1\over2}\left\{\Theta(t_{\rm i}-u)\int_{t_{\rm i}-u}^{v-t_{\rm i}}
+\Theta(u-t_{\rm i})\int_{u-t_{\rm i}}^{v-t_{\rm i}}\right\}
dxS(t_{\rm i}+x),
\label{eq:gamma-sol}
\end{equation}
where $\Theta(x)$ is Heaviside's step function, and we have imposed a 
boundary condition $\delta_{\gamma}{}'|_{r=0}=0$ so that
the analyticity, or equivalently, the locally Minkowskian nature 
at the origin $r=0$ is guaranteed before the
formation of the spacetime singularity there.
We can also obtain a solution of Eq.(\ref{eq:R-linear}) by
the ordinary procedure to obtain the solutions of the wave equation
in 2-dimensional Minkowski spacetime. 
Here we should note that $r\delta_{R}$ has to vanish at $r=0$ 
before the singularity formation. 
The solution satisfying this boundary condition is given as
\begin{equation}
\delta_{R}(t,r;t_{\rm i})=
{2\over \pi r}\int_{t_{\rm i}}^{t}
d\tau\int_{0}^{\infty}dx S(\tau+x)
\int_{0}^{\infty}{dk\over k}
\sin(kr)\sin(kx)\sin\left\{k(t-\tau)\right\}.
\label{eq:R-sol}
\end{equation}
We would like to express the solution of Eq.(\ref{eq:psi-linear})
in the form
\begin{equation}
\psi(t,r;t_{\rm i})
=\int_{-\infty}^{+\infty}dv_{\rm s} S(v_{\rm s})\Psi(t,r;t_{\rm i},v_{\rm
s}),
\label{eq:psi-sol}
\end{equation}
where by introducing the following function,
\begin{equation}
\Omega(t,r;x,y):={(t-y)^{2}-r^{2}-2(t-y)(x-y)\over 2r(x-y)},
\label{eq:Omega-def}
\end{equation}
$\Psi$ is expressed as
\begin{equation}
\Psi(t,r;t_{\rm i},v_{\rm s})
={1\over \pi}\sqrt{2|t_{\rm i}-v_{\rm s}|\over r}
\int_{\Omega(t,r;t_{\rm i},v_{\rm s})}^{1}{dz\over z+(t-v_{\rm s})/r}
\sqrt{z-\Omega(t,r;t_{\rm i},v_{\rm s}) \over 1-z^{2}}.
\label{eq:Psi-def}
\end{equation}
The detailed derivation of the above equation is given in
Appendix. 
Eq.(\ref{eq:conservation-linear}) can be easily integrated
and the solution satisfying the initial condition (\ref{eq:initial})
is obtained as
\begin{equation}
\delta_{D}
=\psi-\delta_{\gamma}+(t_{\rm i}-t)\left(V{d\ln S \over dv}-2S\right).
\label{eq:conservation-sol}
\end{equation}
The lowest order of the energy flux vector in the vacuum region
is given by
\begin{eqnarray}
\sqrt{-g}J^{t}&=&{r\over 8\pi G}\left({\dot \psi}^{2}+{\psi'}^{2}\right), \\
\sqrt{-g}J^{r}&=&-{r\over 4\pi G}{\dot \psi}\psi'. \label{eq:Jr-linear}
\end{eqnarray}
Hence we need to know the solution for $\psi$ to see the
energy flux at the future null infinity.

As mentioned, the symmetric axis $r=0$ with non-vanishing $D_{B}$ is 
the naked singularity of the background Morgan spacetime and   
will also be the naked singularity of the spacetime considered here. 
If $S(v)$ and $Vd\ln S/dv$ are functions defined in $r\geq0$, 
the solutions of perturbation quantities 
(\ref{eq:gamma-sol})-(\ref{eq:conservation-sol}) 
could be finitely defined in $t\geq t_{\rm i}$ and thus in the causal 
future of the naked singularity. 
This means that we could study the structure of the naked singularity 
defined by Eqs.(\ref{eq:gamma-sol})-(\ref{eq:conservation-sol}). 
However here note that in general, in order to solve the causal 
future of naked singularities as a Cauchy problem, 
we need to impose some boundary conditions at the 
naked singularities. In reality, we have imposed the boundary conditions 
for the perturbation equations (\ref{eq:gamma-linear})-(\ref{eq:psi-linear})
at the naked singularity to 
obtain the solutions in the causal future of the naked singularity; 
the symmetric axis $r=0$ is locally Minkowskian. 

By imposing this boundary 
condition, we constructed the Green's function of each perturbation equation 
and then wrote down the solutions (\ref{eq:gamma-sol})-(\ref{eq:psi-sol}). 
However, after the formation of the background naked singularity, 
it is meaningless to impose the locally Minkowskian nature of the axis $r=0$,  
since the axis is singular. Hence we have to be careful 
to discuss the above solutions in the causal future 
of the naked singularity. 
For this reason, in this paper, we focus on the causal past of the 
Cauchy horizon only. 

\subsection{Thin Shell}

First, we consider a thin shell, that is,
\begin{equation}
S=S_{\rm c}\delta(v), \label{eq:D-shell}
\end{equation}
where $S_{\rm c}$ is a positive constant and $\delta(v)$ is Dirac's delta
function. We set $V=V_{\rm c}=$constant.
Since the spacetime singularity forms
at $(t,r)=(0,0)$, we should choose the initial time
$t_{\rm i}$ negative. The Cauchy horizon, 
which is the boundary of the causal future of the naked
singularity, is given by $u=0$. 
From Eqs.(\ref{eq:DB-def}) and (\ref{eq:D-shell}), we obtain 
\begin{equation}
\gamma_{B}={S_{\rm c}\over V_{\rm c}}\Theta(v), \label{eq:gamma-shell}
\end{equation}
where we have imposed the boundary condition, $\gamma_{B}=0$
at $r=0$, so that 
the symmetric axis $r=0$ is locally Minkowskian 
before the singularity formation.

In this thin shell case, as will be shown below, the high-speed 
approximation breaks down. This means that Echeverria's numerical results for
the gravitational collapse of a thin dust shell can not be recovered
by our method. However, the solution of a thin shell is useful in
constructing solutions for the thick shell case and hence we proceed 
to study this. 

From Eq.(\ref{eq:gamma-sol}),
we obtain the solution for $\delta_{\gamma}$ as
\begin{equation}
\delta_{\gamma}(t,r;t_{\rm i})=-{S_{\rm c}\over2}\left\{\Theta(2t_{\rm i}-u)
-\Theta(v)\right\}.
\end{equation}
From Eq.(\ref{eq:R-sol}), we obtain the solution for $\delta_{R}$
in the region of $t\geq t_{\rm i}$ and of the causal past of the 
Cauchy horizon $u\leq0$ as
\begin{eqnarray}
\delta_{R}(t,r;t_{\rm i})&=&{S_{\rm c} \over 2r}\bigl[
\left(t-t_{\rm i}\right)
\left\{\Theta(v)-\Theta(u)+\Theta(u-2t_{\rm i})-\Theta(v-2t_{i})\right\}
\nonumber \\
&-&{v\over2}\Theta(u-2t_{\rm i})\Theta(v)+{u\over2}\Theta(v-2t_{\rm
i})\Theta(u)
\bigr].
\end{eqnarray}
Substituting Eq.(\ref{eq:D-shell})
into Eq.(\ref{eq:psi-sol}), we obtain
\begin{equation}
\psi_{\rm s}(t,r;t_{\rm i})=S_{\rm c}\Psi(t,r;t_{\rm i},0).
\label{eq:psi-sol-thin}
\end{equation}
In order to see the asymptotic behavior in approaching to the
Cauchy horizon, we define a new variable $\varepsilon$ as
\begin{equation}
t=r(1-\varepsilon).
\end{equation}
Then the function $\Psi$ is expressed as
\begin{equation}
\Psi(t,r; t_{\rm i},0)={1\over \pi}\sqrt{2|t_{\rm i}|\over r}
\int_{0}^{2-\varepsilon_{2}}
{\sqrt{x}dx\over (x+\varepsilon_{1})\sqrt{(x+\varepsilon_{2})
(2-\varepsilon_{2}-x)}},
\end{equation}
where we have introduced a new integration variable
$x=z-\Omega(t,r;t_{\rm i},0)$ and
\begin{eqnarray}
\varepsilon_{1}&:=&{r\over |t_{\rm i}|}\varepsilon
\left(1-{\varepsilon\over 2}\right), \\
\varepsilon_{2}&:=&{r\over |t_{\rm i}|}\varepsilon
\left(1+{|t_{\rm i}|\over r}-{\varepsilon\over 2}\right).
\end{eqnarray}
The limit of $\varepsilon\rightarrow0_{+}$ with $r$ fixed
corresponds to the limit to approach
the Cauchy horizon from its chronological past.  
We can easily see that $\Psi$ becomes infinite
in the limit of $\varepsilon \rightarrow0_{+}$ with $r$ fixed.
This divergence of $\Psi$ comes from the integral in the neighborhood
of $x=0$. Hence when $0<\varepsilon\ll1$, $\Psi$ is dominated by the
integral from $x=0$ to $x=\delta x$, where $\delta x$ is some
sufficiently small positive constant;
\begin{eqnarray}
\Psi(t,r; t_{\rm i},0)&\longrightarrow&
{1\over \pi}\sqrt{2|t_{\rm i}|\over r}\int_{0}^{\delta x}
{x^{1/2}dx\over (x+\varepsilon_{1})\sqrt{x+\varepsilon_{2}}}
\left\{{1\over\sqrt{2}}
+O(\varepsilon,\delta x)\right\} \nonumber \\
&=&{2\over \pi}\sqrt{|t_{\rm i}|\over r}\int_{0}^{\sqrt{\delta x}}
{y^{2}dy\over (y^{2}+\varepsilon_{1})\sqrt{y^{2}+\varepsilon_{2}}}
+O(\varepsilon,\delta x) \nonumber\\
&=& -{1\over \pi}\sqrt{|t_{\rm i}|\over r}
\ln \varepsilon+O(1). \label{eq:Psi-asymptotic-1}
\end{eqnarray}
Thus the asymptotic behavior in the limit to approach the Cauchy
horizon is given as
\begin{equation}
\psi(t,r;t_{\rm i})\longrightarrow -{S_{\rm c}\over \pi}
\sqrt{|t_{\rm i}|\over r}
\ln|t-r|.
\end{equation}
Using the above result and from Eq.(\ref{eq:Jr-linear}), 
we find that the energy flux behaves as
\begin{equation}
\sqrt{-g}J^{r}\longrightarrow {S_{\rm c}{}^{2}|t_{\rm i}|
\over 4G\pi^{3} (t-r)^{2}}
\end{equation}
The energy $\Delta E$
radiated from the neighborhood of the naked singularity is then given by
\begin{equation}
\Delta E=2\pi\int^{t}\sqrt{-g}J^{r}dt
\longrightarrow {S_{\rm c}{}^{2}|t_{\rm i}| \over 4G \pi^{3} |t-r|}.
\end{equation}
Hence the radiated energy diverges on the Cauchy horizon.
This result means that the high-speed approximation scheme breaks 
down there. Using Eq.(\ref{eq:conservation-sol}), we obtain
\begin{equation}
\delta_{D}
=\psi-\delta_{\gamma}+(t_{\rm i}-t)
\left\{V_{\rm c}{d\ln \delta(v)\over dv}-2S_{\rm c}\delta(v)\right\}.
\label{eq:conservation-sol-3}
\end{equation}
The above equation shows that $\delta_{D}$ is infinite
and this means that the high-speed approximation 
analysis is not applicable to the
thin shell case.

\subsection{Thick Shell Case}

Let us investigate the gravitational collapse of a thick shell with 
smooth $\gamma_{B}$. Here we consider the following
$S$ and $V$;
\begin{eqnarray}
S&=&{\cal S}_{\rm c} v^{2}
\left(v_{\rm w}-v\right)^{2}
\Theta(v)\Theta(v_{\rm w}-v), \label{eq:S-value} \\
V&=&{\cal V}_{\rm c}v\left(v_{\rm w}-v\right), \label{eq:V-value}
\end{eqnarray}
where $v_{\rm w}$, ${\cal S}_{\rm c}$ and ${\cal V}_{\rm c}$
are positive constants.
Also in this case, the Cauchy horizon is located at $u=0$.
Since $\delta_{\gamma}$ and $\delta_{R}$ are finite
in the causal past of the Cauchy horizon in the thin shell case, 
it is trivial that these are also finite in the thick shell case.
Thus hereafter we focus on $\psi$ and $\delta_{D}$. 

Using the asymptotic solution (\ref{eq:Psi-asymptotic-1}) in the thin shell
case, we can easily estimate the asymptotic behavior of
Eq.(\ref{eq:psi-sol}) also in the thick shell case.
Here we focus on the region of sufficiently large $r$ and hence 
we assume that $v_{\rm w}/r$ is much smaller than unity. 
In the integrand of Eq.(\ref{eq:Psi-def}), 
$0<v_{\rm s}/r\leq v_{\rm w}/r \ll 1$ holds by 
the assumption and thus by the same procedure to derive
Eq.(\ref{eq:Psi-asymptotic-1}), in the limit to the Cauchy horizon 
$\varepsilon\rightarrow0_{+}$, we obtain
\begin{eqnarray}
\psi(t,r)&\sim&-{{\cal S}_{\rm c}\over\pi}\sqrt{|t_{\rm i}|\over r}
\int_{0}^{v_{\rm w}}dv_{\rm s}v_{\rm s}{}^{2}\left(v_{\rm w}-v_{\rm
s}\right)^{2}
\ln\left(\varepsilon+{v_{\rm s}\over r}\right)
\longrightarrow
-{{\cal S}_{\rm c}v_{\rm w}{}^{5}\over 30\pi}\sqrt{|t_{\rm i}|\over r}
\ln\left({v_{\rm w}\over r}\right),
\label{eq:psi-asymptotic-2}
\end{eqnarray}
where we have also assumed $v_{\rm w}/|t_{\rm i}|\ll1$. 
Thus as expected, $\psi$ does not diverge even on the Cauchy horizon $u=0$.
The energy flux is also finite in the same limit; we can easily
verify that the asymptotic value in the limit to the Cauchy
horizon for sufficiently large $r$ is 
\begin{equation}
\sqrt{-g}J^{r}\longrightarrow
{25|t_{\rm i}|(v_{\rm w}{}^{4}{\cal S}_{\rm c})^{2}\over 576\pi^{3}G}
\label{eq:C-energy-flux}
\end{equation}
Using Eqs.(\ref{eq:S-value}) and (\ref{eq:V-value}),
Eq.(\ref{eq:conservation-sol}) becomes as
\begin{equation}
\delta_{D}=\psi-\delta_{\gamma}
+2(t_{\rm i}-t)\left\{{\cal V}_{\rm c}(v_{\rm w}-2v)
-{\cal S}_{\rm c} v^{2}(v_{\rm w}-v)^{2}\right\}
\Theta(v)\Theta(v_{\rm w}-v).
\end{equation}
From the above equation, we find that $\delta_{D}$ is finite
in the causal past of the Cauchy horizon. Together with the result of
$\psi$, we may say that the solutions of the high-speed approximation
can describe the asymptotic behavior of the gravitational collapse of
the thick shell. 

In the limit of $v_{\rm w}\rightarrow0$ with
${\cal S}_{\rm c}v_{\rm w}{}^{5}$ fixed, this model approaches to the
infinitely thin shell with finite line density. 
Eqs.(\ref{eq:psi-asymptotic-2}) and
(\ref{eq:C-energy-flux}) show that in this limit, $\psi$ and
correspondingly energy flux $J^{r}$ become infinite. This result is
consistent with the preceding analysis of the thin shell case. 
The width $v_{\rm w}$ of the shell corresponds to the time difference of the
singularity formation; the innermost dust fluid element first forms
the spacetime singularity at $t=0$ while the singularity formation time
of the outermost fluid element is $t=v_{\rm w}$. 
Hence, almost simultaneous naked singularity 
formation will lead to huge emission of gravitational radiation.

\section{Summary and Discussion}

We analytically constructed approximate solutions for cylindrically
symmetric gravitational collapse of a dust shell with finite width  
by the high-speed approximation. 
This approximation scheme could treat the last stage of the
gravitational collapse at which the collapsing speed is very large due
to the gravitational acceleration. 
We found that thinner width
of the shell leads to larger amount of gravitational radiation.
Hence, if the dust fluid of finite mass collapses into the spacetime
singularity almost simultaneously, a gravitational wave burst might occur
and this is consistent with Echeverria's numerical results. 
Our results also imply that the trajectory of the 
cylindrical thick shell does not agree with null in the limit to 
the Cauchy horizon and the gravitational emission does not cease there.
On the other hand, Echeverria's speculation 
by using self-similar asymptotic solutions is not so in   
the case of an infinitely thin shell. 
Here it should be noted that as shown in Sec.III, 
the high-speed approximation breaks down 
in the infinitely thin shell case and hence our results cannot be
directly compared with Echeverria's results. 
Therefore it is a future work to reveal the physical 
reason of this difference between the infinitely thin 
shell case and our present one.

Our result agrees with Chiba's numerical 
analysis. In these simulations, the gravitational collapse
of an infinitesimal portion at $r=0$ first forms a massless
singularity and then the dust fluid in the outer region will accrete on 
this singularity. 
In other words, the spacetime singularity is gradually formed in 
Chiba's numerical solutions and hence
the gravitational waves are not emitted so much in the 
causal past of the Cauchy horizon. This situation could be
realized also in the gravitational collapse of the collisionless
gas spheroid in the numerical simulations by Shapiro and
Teukolsky\cite{Ref:Spindle}; also in their case, appreciable
gravitational waves are not generated. 

From our analytic and Echeverria's numerical results,  
the cylindrical dust collapse could be a strong source of gravitational 
radiation if a finite amount of dust fluid almost simultaneously collapses 
to form a naked singularity. Of course, the cylindrical distribution
is not realistic and hence our results do not directly lead to 
a conclusion that formation processes of naked singularities 
are promising targets of the gravitational wave astronomy. 
Further studies are necessary to obtain more definite 
implications. However we would like to note that recently a possibility 
of the naked singularity formation 
process as a candidate of gamma-ray burst has been 
also discussed\cite{Ref:HIN00,Ref:JDM}. 
Hence if  singularities formed in our universe are 
globally naked, those might have astronomical significance in 
addition to their importance as a laboratory of new physics. 

\section*{Acknowledgements}
We are grateful to H.~Ishihara and colleagues in the astrophysics and
gravity group of Osaka City University for useful and helpful 
discussion and criticism. KN thanks S.~A.~Hayward for his important
comment in JGRG workshop in November 2002. 
YM thanks the Yukawa Memorial Foundation for its support.

\appendix
\section{Formal Solution of $\psi$} 

The Green function $G_{3}$ of the wave equation in 3-dimensional
Minkowski spacetime is determined by
\begin{equation}
\left(\partial_{t}^{2}-\partial_{r}^{2}-{1\over r}\partial_{r}
-{1\over r^{2}}\partial_{\varphi}^{2}\right)G_{3}(t,r,\varphi;\tau,x,\phi)
={1\over r}\delta(t-\tau)\delta(r-x)\delta(\varphi-\phi),
\end{equation}
where $r$ and $x$ are radial coordinate $[0,+\infty)$, and
$\varphi$ and $\phi$ are angular coordinate $[0,2\pi)$.
The retarded solution of the above equation is obtained as
\begin{eqnarray}
G_{3}(t,r,\varphi;\tau,x,\phi)
={\Theta(\Delta t-\Delta r) \over 2\pi
\sqrt{(\Delta t)^{2}-(\Delta r)^{2}}},
\label{eq:Green-func}
\end{eqnarray}
where
\begin{eqnarray}
\Delta t&:=&t-\tau, \\
\Delta r&:=&\sqrt{r^{2}+x^{2}-2rx\cos(\varphi-\phi)}.
\end{eqnarray}

Using the retarded Green function,
the solution for $\psi$ is formally written in the form
\begin{eqnarray}
\psi(t,r)&:=&\int_{t_{\rm i}}^{t}d\tau
\int_{0}^{\infty}dx \int_{0}^{2\pi}d\phi
S(\tau+x) G_{3}(t,r,0;\tau,x,\phi) \nonumber \\
&=&\int_{-\infty}^{+\infty}dv_{\rm s}S(v_{\rm s})
\int_{t_{\rm i}}^{t+0}d\tau
\int_{0}^{\infty}dx \int_{0}^{2\pi}d\phi
\delta(\tau+x-v_{\rm s})G_{3}(t,r,0;\tau,x,\phi),
\label{eq:psi-sol-formal}
\end{eqnarray}
Comparing Eq.(\ref{eq:psi-sol}) with Eq.(\ref{eq:psi-sol-formal}),
we find that
\begin{equation}
\Psi(t,r;t_{\rm i},v_{\rm s})
=\int_{t_{\rm i}}^{t}d\tau
\int_{0}^{\infty}dx \int_{0}^{2\pi}d\phi
\delta(\tau+x-v_{\rm s})G_{3}(t,r,0;\tau,x,\phi).
\end{equation}

Using Eq.(\ref{eq:Green-func}), for the region
$v>v_{\rm s}$ and $u<v_{\rm s}$,  $\Psi$ is calculated as
\begin{eqnarray}
\Psi(t;r)&=&{1\over 2\pi}
\int_{t_{\rm i}}^{(t-v_{\rm s}-r)/2+v_{\rm s}}d\tau
\int_{-\arccos\Omega(t,r;\tau,v_{\rm s})}^{+\arccos\Omega(t,r;\tau,v_{\rm
s})}
d\phi \nonumber \\
&&~~~~~~~~~~~~~~\times
{1\over \sqrt{(\tau-t)^{2}
-\{(\tau-v_{\rm s})^{2}+2r(\tau-v_{\rm s})\cos\phi+r^{2}\}}} \nonumber \\
&& \nonumber \\
&=&{1\over \pi}\int_{\Omega(t,r;t_{\rm i},v_{\rm s})}^{1}{dz\over
\sqrt{1-z^{2}}}
\int_{t_{\rm i}-v_{\rm s}}^{(t-v_{\rm s})^{2}-r^{2}\over 2(rz+t-v_{\rm s})}
{d\tau_{\rm s}\over \sqrt{(t-v_{\rm s})^{2}-r^{2}
-2(rz+t-v_{\rm s})\tau_{\rm s}}} \nonumber \\
&& \nonumber \\
&=&{1\over \pi}\sqrt{2|t_{\rm i}-v_{\rm s}| \over r}
\int_{\Omega(t,r;t_{\rm i},v_{\rm s})}^{1}{dz\over z+(t-v_{\rm s})/r}
\sqrt{z-\Omega(t,r;t_{\rm i},v_{\rm s}) \over 1-z^{2}},
\label{eq:Psi-sol}
\end{eqnarray}
where $\Omega$ is defined by Eq.(\ref{eq:Omega-def}).
In the second equality of Eq.(\ref{eq:Psi-sol}), after introducing
new integration variables $\tau_{\rm s}=\tau-v_{\rm s}$ and
$z=\cos{\phi}$, we  change the order of the integrations with respect to
$\tau_{\rm s}$ and $z$. Then in the final equality, the integration
with respect to $\tau_{\rm s}$ is performed.

\newpage

\begin{figure}
\rotatebox{-90}{\resizebox{10cm}{!}{\includegraphics{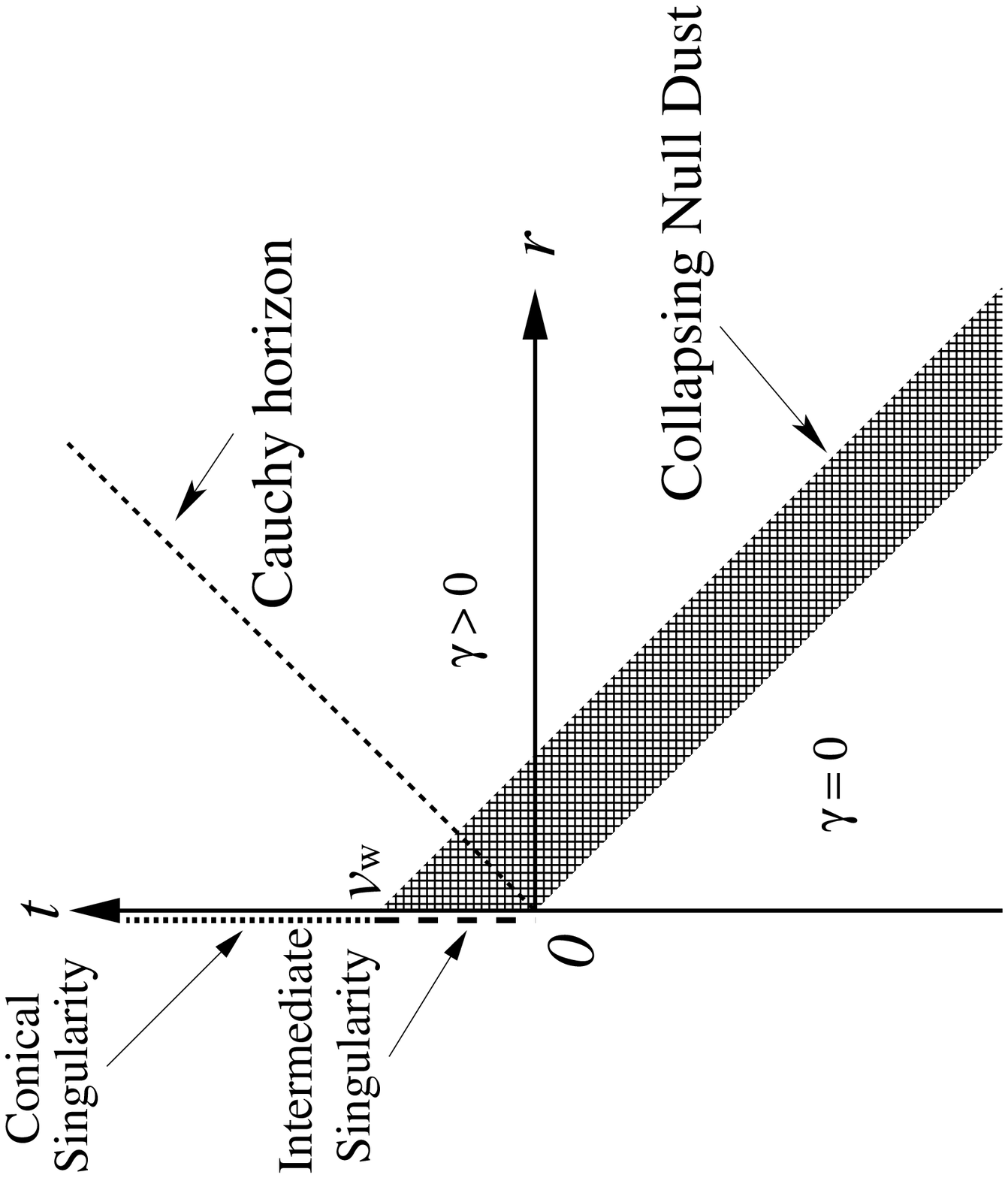}}}
\caption{Morgan's cylindrical null dust solution. 
There is the null dust in the shaded region; $D(v)>0$ for 
$0<v<v_{\rm w}$. The dashed line on $t$-axis corresponds to 
the intermediate singularity at which an observer suffers infinite 
tidal force although any scalar polynomials of the Riemann tensor 
do not vanish.  On the other hand, the dotted line on $t$-axis 
is the conical singularity. The Cauchy horizon is short dashed line of $t=r$. 
}
\label{fg:M-detector}
\end{figure}

\end{document}